\documentstyle[12pt,epsfig]{article}
\newlength{\dinwidth}
\newlength{\dinmargin}
\newcommand{\ba}{\begin{array}}
\newcommand{\ea}{\end{array}}
\newcommand{\be}{\begin{equation}}
\newcommand{\ee}{\end{equation}}
\newcommand{\bea}{\begin{eqnarray}}
\newcommand{\eea}{\end{eqnarray}}
\newcommand{\gsim}{\mathrel{\mathop{\kern 0pt \rlap
  {\raise.2ex\hbox{$>$}}} \lower.9ex\hbox{\kern-.190em $\sim$}}}

\def\ben{\begin{equation}}
\def\een{\end{equation}}
\def\bea{\begin{eqnarray}}
\def\eea{\end{eqnarray}}
\def\nn{\nonumber}

{\rm

\def\cH{{\cal H}}

\begin{document}

\thispagestyle{empty}
\addtocounter{page}{-1}
\vskip-0.35cm
\begin{flushright}
UK/07-01 \\
{\tt hep-th/yymmxxx}
\end{flushright}
\vspace*{0.2cm}
\centerline{\Large \bf Open String Descriptions of}
\centerline{\Large \bf Space-like Singularities
in}
\centerline{\Large \bf Two Dimensional String Theory}
\vspace*{1.0cm} \centerline{{\bf Sumit R. Das} and {\bf Luiz H. Santos}}
\vspace*{0.7cm}
\centerline{\it Department of Physics and Astronomy,}
\vspace*{0.2cm}
\centerline{\it University of Kentucky, Lexington, KY 40506 \rm USA}
\vspace{0.7cm}
\centerline{\tt das@pa.uky.edu,~~luiz.santos@uky.edu}

\vspace*{0.8cm}
\centerline{\bf Abstract}
\vspace*{0.3cm}
\noindent The matrix model formulation of two dimensional string
theory has been shown to admit time dependent classical solutions
whose closed string duals are geodesically incomplete
space-times with space-like boundaries. We investigate some aspects of
the dynamics of fermions in one such background. We show that even
though the background solution appears pathological, the time
evolution of the system is smooth in terms of open string degrees of
freedom, viz. the fermions. In particular, an initial state of
fermions evolves smoothly into a well defined final state over an
infinite open string time interval, while the time perceived by closed
strings appears to end abruptly. We outline a method of calculating
fermion correlators exactly using symmetry properties. The result for
the two point function is consistent with the semiclassical picture.

\vspace*{0.5cm}

\newpage

\section{Introduction}

Recently several examples of space-like and null singularities in
string theory have been analyzed using holographic dual
formulations. These include backgrounds in the matrix model
formulation of two dimensional string theory which
lead to spacelike boundaries in the closed string 
interpretation\cite{Das:2004aq}, ten
dimensional backgrounds which admit a Matrix Theory type formulation
\cite{Craps:2005wd,li0506,Das:2005vd,robbinssethi0509}, and
deformations of $AdS$ space-times which admit a dual gauge theory
description \cite{Das:2006pw,Chu:2006pa}. In all these examples, 
(which build on a large body of earlier work on strings on time
dependent background \cite{refer}), the
low energy space-time description breaks down at the singularity as
expected. However in each case the open string dual appears to be well
defined. This realizes a long held belief that near singularities the
usual notions of space and time have to be abandoned and replaced by
some more fundamental structure : in these cases this structure is
provided by the open string dual.

The examples in two dimensional string theory are particularly
significant in this respect since the dual matrix model reduces to a model of
free fermions in an external inverted harmonic potential 
and therefore in some sense solvable. In recent years it
has been realized that the matrix model / string theory connection is in
fact open-closed duality just like the AdS/CFT correspondence
\cite{McGreevy:2003kb}. The matrix degrees of freedom - and hence the
fermionic eigenvalues - are the open strings, while the closed string
description is most conveniently provided by collective field theory
\cite{Jevicki:1979mb, Das:1990ka}. The collective field is a
bosonization of the fermion field. Nontrivial time dependent classical
solutions of the collective field theory correspond to time dependent
fermi surfaces. Such backgrounds have been studied for quite a while
\cite{Minic:1991rk}. In \cite{Karczmarek:2003pv, Karczmarek:2004ph,
Das:2004hw} a class of such backgrounds were studied as toy models of
cosmology.

In \cite{Das:2004aq} it was shown that there is a class of backgrounds
whose closed string intepretation involve space-times with a
space-like boundary. In these latter backgrounds, the time of the
matrix model - which we call the open string time - runs over the full
range $-\infty < t < \infty$. However, the time in terms of which the
collective field fluctuations define a relativistic theory - which we
call the closed string time - stops abruptly, thus forming a
space-like ${\cal I}^+$. This ${\cal I}^+$ is however not an
asymptotic region since the coupling is non-vanishing, though there is
a part of ${\cal I}^+$ where the coupling becomes arbitarily weak.

At first sight, it appears that the space-like singularity is caused  
by the fact that in this background the 
eigenvalue space shrinks to zero at late times so that closed strings have
no space to propagate. The same fact might also suggest 
that there is no well defined scattering problem. The analysis of 
\cite{Das:2004aq} however showed that the "space" on which the closed
string modes propagate is related to the space of eigenvalues in a 
non-trivial time dependent way. As a result, this {\em closed string space}
is still of infinite extent at arbitrarily late times and there is 
indeed a well posed scattering problem.  
This conclusion is based on the quadratic part of the action
for collective field fluctuations. Since a large part of ${\cal I}^+$
is strongly coupled, one might worry that the effects of nonzero coupling
may singinificantly modify this picture.

In this paper we re-visit this model from the point of view of the
fermionic theory so that the couplings of the collective field
are treated in an exact fashion.  We first study the classical evolution of small
ripples on the fermi sea and derive an equation which determines the
shape of the ripple at late times in terms of the initial state,
analogous to the scattering equation for perturbations around the
ground state \cite{Polchinski:1991uq, Natsuume:1994sp}. We find that
initial pulses which are approximately localized at
early or late retarded time scatter into final pulses which are 
localized on the (spacelike) ${\cal I}^+$ at large values of the closed
string spatial coordinate $q$ - as expected from linearized collective
field theory. This is consistent, since for large $q$ the coupling is
weak. For small retarded times, the scattered
pulse is still localized, although 
its location is shifted compared to the expectations of
free collective theory.  

We then establish a general relation between
{\em exact} fermion correlators in the time dependent background and
those in the ground state. This enables a calculation of these
correlators in terms of those obtained in \cite{Moore:1991sf}. We
illustrate this by an exact calculation of the eigenvalue density. The
result again shows that the picture based on collective field theory
is reliable in the expected regime.

The present paper deals with one particular background in 
\cite{Das:2004aq}. The methods and results are, however,
expected to be similar for the other backgrounds discussed
in that paper.

In Section 2, we review the background solution and the behavior
of collective field fluctuations at the linearized level described
in \cite{Das:2004aq}. Section 3 describes some aspects of the
{\em classical} dynamics of small ripples on the fermi sea using the
exact fermion equations of motion, following 
\cite{Polchinski:1991uq, Natsuume:1994sp}. This is used in Section 4
to derive a scattering equation for such ripples riding on the
time dependent background. In Section 5 we obtain profiles of
the scattered pulse for an initial gaussian pulse by numerically
investigating the scattering equations. In section 6 we outline a
general method for calculating {\em exact} correlation functions of
fermions in such backgrounds, utilizing properties of $W_\infty$
symmetry and obtain an explicit expression for the eigenvalue 
expectation value.

\section{$c=1$ Matrix Model and 2d String Theory}

In this section we review the way (approximate) relativistic space-time
appears in the $c=1$ matrix model \cite{ceqone}. 

The dynamical variable of the model is a single $N \times N$ matrix
$M_{ij}(t)$ 
and there is a constraint which restricts the states to be singlets.
In the singlet sector, and in the double scaling limit \cite{ceqone},
matrix quantum
mechanics reduces to a theory of an infinite number of 
fermions with the single particle
hamiltonian given by
\ben
H = {1\over 2}[p^2 - x^2]
\label{one}
\een
where we have adopted conventions in which the string scale
$\alpha^\prime = 1$ for the bosonic string and $\alpha^\prime =
{1\over 2}$ in the Type 0B string. The fermi energy in this rescaled
problem will be denoted by $-\mu$.

In the classical limit, the system is equivalent to an incompressible
fermi fluid in phase space. The ground state is the static fermi
profile
\ben
(x-p)(x+p)=2\mu
\label{two}
\een

The dual closed string theory is best obtained by rewriting the theory
in terms of the collective field $\phi (x,t)$ which is defined as the
density of eigenvalues of the original matrix. 
\ben
\partial_x \phi (x,t) = {1\over N}{\rm Tr}~\delta (M(t)-x\cdot I)
\een
At the classical level
the action of the collective field is given by
\ben
S = N^2\int dxdt~\left[{1\over 2}{(\partial_t \phi)^2 \over (\partial_x \phi)}
  - {\pi^2 \over 6} (\partial_x \phi)^3 - (\mu - {1\over
    2}x^2)\partial_x \phi \right]
\label{aone}
\een
This is of course a theory in $1+1$ dimensions, the spatial dimension
arising out of the space of eigenvalues.

Fermi seas with quadratic profiles appear as classical solutions to
collective field theory in the appropriate limit \footnote{
Quadratic profiles are those fermi surfaces $f(x,p)=0$ where $p$
appears at most quadratically. Profiles which are not
quadratic do not correspond to {\em classical} solutions of
collective field theory \cite{Dhar:1992cs}. Rather they are
highly quantum states of the collective theory in which 
quantum dispersions do not vanish in the classical limit
\cite{Das:1995gd}.}.
The space-time which is generated may be obtained 
by looking at the dynamics of fluctuations of the collective field
around the classical solution. Expanding
around an {\em arbitrary} classical solution $\phi_0 (x,t)$
\ben
\phi(x,t) = \phi_0 (x,t) + {1\over N}\varphi (x,t)
\label{asix}
\een
The action for these fluctuations at the quadratic level may be
written as
\ben
S^{(2)}_\varphi = {1\over 2} \int dt dx~{\sqrt{g}}g^{\mu\nu}\partial_\mu
\varphi \partial_\nu \varphi
\label{aseven}
\een
where $\mu,\nu = t,x$. The line element determined by 
$g_{\mu\nu}$ is conformal to 
\ben
ds^2 = -dt^2 + {(dx + {\partial_t \phi_0 \over \partial_x \phi_0}dt)^2 
\over (\pi \partial_x \phi_0)^2}
\label{aeight}
\een
Therefore, {\em regardless of the classical solution}, the spectrum is
always a massless scalar in one space dimension given by $x$.
The metric can be determined only upto a conformal factor. However, as
we will see below, the global properties of the space-time can be
determined from the nature of the classical solution.

The classical 
interaction hamiltonian is purely cubic when expressed in terms of the
fluctuation field $\varphi$ and its canonically conjugate momentum
$\Pi_\varphi$, 
\ben
H^{(3)}_\varphi = \int dx [{1\over 2}\Pi_\varphi^2 \partial_x\varphi + {\pi^2
    \over 6} (\partial_x\varphi)^3]
\label{anine}
\een

\subsection{The Ground State and its fluctuations}

The ground state (\ref{two}) is a quadratic profile and
the classsical solution is
\ben
\partial_x \phi_0 = {1\over \pi}
	{\sqrt{x^2-2\mu}}~~~~~\partial_t\phi_0 = 0 
\label{atwo}
\een

Around the ground state, the metric (\ref{aeight}) is given by
\ben
ds^2 = -dt^2 + {dx^2 \over x^2 -2\mu}
\label{aten}
\een
The perturbative fluctuations live in the region $|x| > {\sqrt{2\mu}}$
and the field $\varphi$ satisfies Dirichlet boundary condition at the
``mirrors'' given by $x = \pm {\sqrt{2\mu}}$. The fields on the
``left'' and ``right'' side are decoupled at the perturbative level. 
The physics of these fields
is made transparent by choosing Minkowskian coordinates
$(\sigma,\tau)$ which in this case are
\ben
t = \tau ~~~~~~~~~ x = \pm {\sqrt{2\mu}} \cosh \sigma
\label{aeleven}
\een
In these coordinates 
\ben
ds^2 = -d\tau^2 + d\sigma^2
\een

\begin{figure}[ht]
\centerline{\epsfxsize=1.5in
\epsfysize=2.0in
   {\epsffile{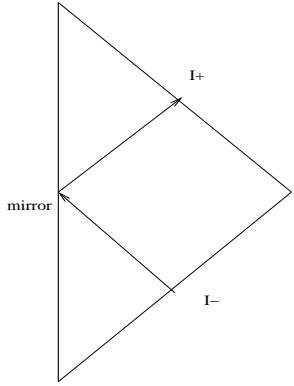}}
}
\caption{Penrose diagram of space-time produced by ground state
  solution showing an incoming ray getting reflected at the mirror}

\label{penrose_ground}
\end{figure}

The field $\varphi$ may be now thought of being made of {\em two} fields,
$\varphi_{S,A}(x,t)$ each of which live in the region $x > 0$
\ben
\varphi_{S,A} (x,t) = {1\over 2}[\varphi (x,t) \pm \varphi(-x,t)]
\label{atwelve}
\een
In terms of the Minkowskian coordinates, solutions to the
linearized equations are plane waves $\varphi_{S,A} \sim
e^{-i\omega(t\pm\sigma)} \varphi_{S,A} (\omega)$ and these fourier
components are related to the two spacetime fields - the tachyon $T$
and the axion $C$ - which appear in the
standard formulation of Type 0B string theory 
\cite{Takayanagi:2003sm} :
\bea
T(\omega) & = & (\pi/2)^{-i\omega/8}~{\Gamma(i\omega/2) \over
\Gamma(-i\omega/2)}~\varphi_S (\omega) \nn \\
C(\omega) & = & (\pi/2)^{-i\omega/8}~{\Gamma((1+i\omega)/2) \over
\Gamma((1-i\omega)/2)}~\varphi_A (\omega)
\label{athirteen}
\eea
In any case, the space-time generated is quite simple. The Penrose
diagram is that of two dimensional Minkowski space with a mirror at
$\sigma = 0$, as shown in Figure (\ref{penrose_ground}). 
The fluctuations are massless particles which come in
from ${\cal I}^-_{L,R}$, get reflected at the mirror, and arrive at
${\cal I}^+_{L,R}$. 

Recall that we are working in string units. The transforms 
(\ref{athirteen})
imply that the position space fields are related by a transform which
is non-local at the string scale. 
Therefore points on the Penrose diagram should be thought 
of as smeared over the string scale. 
But then, this should be true of any Penrose diagram drawn in a string theory.

In terms of the Minkowskian coordinates the
interaction hamiltonian becomes
\ben
H_3 = \int d\sigma{1\over 2\sinh^2\sigma}[{1\over
    2}{\tilde{\Pi_\varphi}}^2\partial_\sigma\varphi + {\pi^2 \over
    6}(\partial_\sigma\varphi)^3] 
\label{bone}
\een
The canonically conjugate momentum  ${\tilde{\Pi_\varphi}}$ satisfies
the standard commutator $ \left[ \varphi (\sigma), {\tilde{\Pi_\varphi}}
(\sigma^\prime) \right] = i \delta (\sigma - \sigma^\prime)$. 
The interactions therefore vanish at $\sigma = \infty$ and
are strong at $\sigma = 0$ - this gives rise to a non-trivial wall
S-matrix. 

In the Type 0B string theory interpretation, the Penrose diagram has
to be folded across the center. 

As emphasized above, the metric is determined 
only upto a conformal transformation. 
So long as the conformal transformation is non-singular, 
this is sufficient to draw Penrose diagrams. 
A conformal transformation would, however, mix up the 
space $\sigma$  and time $\tau$ and it would appear that 
this leads to an ambiguity. 
The special property of the space and time coordinates 
defined above is that 
the {\em interaction Hamiltonian is time-independent} 
with this choice and a conformal transformation would destroy 
this property. This makes the physics transparent and easy to 
compare with string theory results. Of course a 
different coordinatization with a time dependent 
Hamiltonian is physically equivalent and should 
be compared with the string theory results in an appropriately chosen gauge.

\subsection{Time dependent fermi surfaces}

The theory has an infinite number of symmetries, the
$W_\infty$ symmetries . At the classical level,
the generators of these symmetries in fermion phase space are given by
\ben
W_{rs} = e^{(r-s)t}(x-p)^r (x+p)^s
\een
For $r \neq s$ these generators do not commute with the hamiltonian.
Therefore starting with the ground state, one can obtain exact time
dependent solutions by the action of these generators \cite{Das:2004hw}.
We will be
particularly interested in generators $W_{r0}$ and $W_{0r}$. 
These generate the transformations with parameters $\lambda_\pm$
\ben
(x \pm p) \rightarrow (x\pm p) + \lambda_\pm e^{\pm rt}(x\mp p)^{r-1}
\een
and leads to the following fermi surfaces
\ben
x^2-p^2 + \lambda_-~e^{-rt}(x+p)^r + \lambda_+ ~e^{rt} (x-p)^r +
\lambda_+\lambda_- (x^2-p^2)^{r-1} = 2\mu~.
\label{three}
\een
Formally, the state of the fermion system 
is related to the
ground state $|\mu\rangle$ by 
\ben
|\lambda\rangle = {\rm exp}[i\lambda W]|\mu\rangle ~,
\een
where $W$ denotes the $W_\infty$ charge which generates this solution.
However this state is not normalizable and therefore not contained in the
Hilbert space of the model. Rather, this corresponds to
a deformation of the hamiltonian of the theory to 
\ben
{H}^\prime = e^{-i\lambda W}{H}e^{i\lambda W} ~.
\label{vthree}
\een

\subsection{The closing hyperbola solution}

\begin{figure}[ht]
\centerline{\epsfxsize=3.0in
\epsfysize=3.0in
   {\epsffile{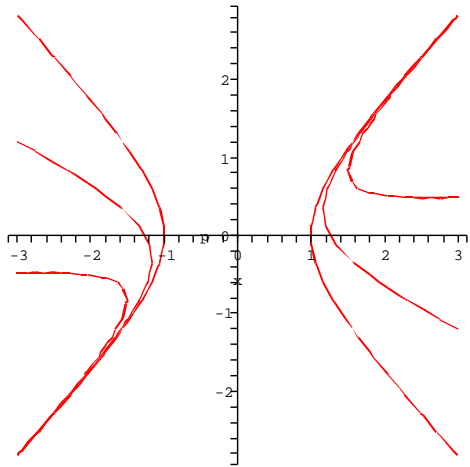}}
}
\caption{The closing hyperbola solution. At late times the hyperbola
  closes on into itself draining all the fermions}
\label{r2fermi1}
\end{figure}

In this paper we will concentrate on the "closing hyperbola solution"
found in \cite{Das:2004aq}. This is the solution generated by the
action of $W_{20}$, i.e. with $\lambda_-=0$ and $\lambda_+ < 0$.
In this case we can choose the origin of time and choose 
$\lambda_+ = -1$. Furthermore $x$ and $p$ may be rescaled to set
$\mu = 1/2$. In the rest of the paper we will stick to these choices.
The classical collective field is then
\ben
\partial_x \phi_0  =  {1\over \pi(1 + e^{2t})}{\sqrt{x^2-(1+e^{2t})}}
~~~~~~
\partial_t \phi_0  =  - {x e^{2t} \over 1+e^{2t}}~\partial_x \varphi_0
\label{afive}
\een
This will be called the ``closing hyperbola'' solution shown and
explained in Figure (\ref{r2fermi1}).

The (approximately) relativistic space-time perceived by the fluctuations
around this classical solution can be once again best seen in Minkowskian
coordinates $q, \tau$ in terms of which the quadratic action is
\ben
S^{(2)} = \int dq \int d\tau [(\partial_\tau \varphi)^2 - (\partial_q \varphi)^2]
\een
and the interactions are independent of $\tau$.
These coordinates are related to the original coordinates $(t,x)$ by the relations 
\ben
x = -\frac{\cosh q}{\sqrt{1-e^{2\tau}}}~~~~~~~~e^t = \frac{e^\tau}
{\sqrt{1- e^{2\tau}}}
\label{bfive}
\een
We have restricted our attention to one side of the inverted harmonic
potential ($x < 0$) and the range of $q$ can be chosen to be
\ben
0 \leq q \leq \infty
\een
To describe the other side we need to have another patch of the $(q,\tau)$
coordinates. In the rest of the paper we will restrict to one side since
for this background the physics of the other side is identical \footnote{This
is not adequate for some of the other solutions described in \cite{Das:2004aq}.}.
The interaction hamiltonian is then given by 
\ben
H_3 = \int dq{1\over 2\sinh^2 q}[{1\over
    2}{\tilde{\Pi_\varphi}}^2\partial_q\varphi + {\pi^2 \over
    6}(\partial_q\varphi)^3] 
\een 

The equations (\ref{bfive}) immediately show
that as $-\infty < t < \infty$ the time $\tau$
has the range $ -\infty < \tau < 0$. Since the dynamics of the matrix
model ends at $t = \infty$ the resulting space-time appears to be
geodesically incomplete with a space-like boundary ${\cal I}^+$ at $\tau = 0$.
This boundary is, however, not an asymptotic region since the coupling is
generally non-vanishing here except for $q \rightarrow \infty$.

\begin{figure}[ht]
\centerline{\epsfxsize=2.0in
\epsfysize=2.0in
   {\epsffile{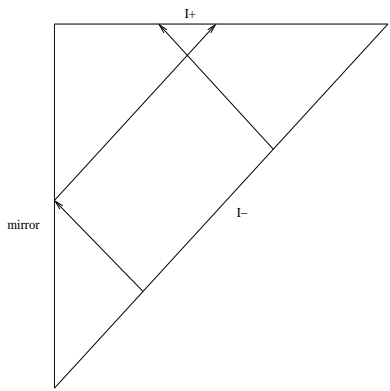}}
}
\caption{Penrose diagram for small fluctuations around 
the closing hyperbola solution showing
  two classes of null rays}
\label{penrose_closing}
\end{figure}

The edge of the fermi sea is at $q=0$, which forms a time-like
reflecting boundary.
If we ignore the couplings of the collective field theory,
fluctuations coming in from ${\cal I}^-$
along $q_+ = q + \tau = \tau_0$ will get reflected by the
mirror at $q = 0$ so long as $\tau_0 < 0$ and hit the space-like
boundary ${\cal I}^+$ at $q = -\tau_0$. For $\tau_0 > 0$ this
ray cannot reach the mirror before time ends - rather it directly hits
the space-like boundary at $q = \tau_0$. The Penrose diagram with these
two classes of rays is shown in Figure (\ref{penrose_closing}). Note,
however that {\em in $x$ space the ray always turns around at some
  point},
as may be seen from the change of variables (\ref{bfive}).

In fact, in the fermion picture the fermi sea gets
drained out and it appears that at $t \rightarrow \infty$ there is
no fermi sea at all, whereas in the bosonic picture there is a 
space-like ${\cal I}^+$. What is happening is that at $t = \infty$,
the entire space-like boundary $\tau = 0$ is at $|x| =\infty$, where 
the coordinate $q$ runs over its full range $0 \leq q \leq \infty$. 
It is therefore clear that the scattering problem has to be
formulated in the $(q,\tau)$ space rather than in the $(x,t)$ space.

Since the bosonic coupling is non-vanishing on the ${\cal I}^+$, one
might worry that interaction effects might substantially modify this
picture. To address these issues we next analyze the dynamics of small ripples
on the fermi sea directly in the fermion picture, thus taking into
acccount the bosonic interactions in an {\em exact} fashion.

\section{Classical dynamics of ripples on the Fermi sea}

We now set up the exact equations for the dynamics of fluctuations around
an arbitary time dependent solution with a quadratic profile,
following the method of \cite{Polchinski:1991uq}. Consider 
a point on the deformed fermi surface labelled by a parameter $\alpha$.
The dynamics of this point is then given by
\ben
x(t)  =  - a(\alpha)~\cosh(t-\alpha),~~~~~~~~~~~~~
p(t)  =  - a(\alpha)~\sinh(t-\alpha) 
\label{neone}
\een
This implies
\ben
x^2(t)-p^2(t)  =  -a^2(\alpha)~~~~~~~~~~~~~
x(t)+p(t)  = - a(\alpha)~e^{t-\alpha}
\label{netwo}
\een
Let this point be at some location $x=x_1$ at some time $t_1$. If this point
returns again to this location at some later time $t_2$, we must have
$p(t_1) = - p(t_2)$. Therefore (\ref{neone})
shows that
\ben
t_1 - \alpha = \alpha - t_2
\label{nethree}
\een
Note that we have assumed that $t_2 > t_1$ and this can happen only if
\ben
|p(t_1)| < |x_1|~~~~~~~~~~~~p(t_1) > 0
\een
If the first inequality is not satisfied the point goes over to
the other side of the potential. If the second inequality is violated,
the point never returns to the same value of $x$ at a later time.

Using the equations of motion to eliminate the parameter $\alpha$, the 
ripple can be described by a function $P(x,t)$. Then the equation
$p(t_1)=-p(t_2)$ provides a {\em scattering equation} for the ripple
\ben
P(x_1,t_1)= - P(x_1,t_2)
\label{scattering1}
\een
Combining (\ref{netwo}) and (\ref{nethree}) we get
\ben
t_2-t_1 = 2(\alpha - t_1) = 2(t_2-\alpha)  =  \log \frac{x_1+
  P(x_1,t_2)}{x_1 - P(x_1,t_2)}
 =  \log \frac{x_1- P(x_1,t_1)}{x_1 + P(x_1,t_1)} 
\label{nefour}
\een

In this paper we will be interested in the evolution of a pulse 
around the fermi surface given by the closing hyperbola solution (\ref{afive}).
This is given by a fermi surface
\ben
x^2-p^2-e^{2t}(x-p)^2=1
\label{nesix}
\een 
which may be solved to obtain the value of momentum at a point
$x$ at time $t$ 
\ben
\bar{P}_\pm(x,t)=\frac{xe^{2t}}{1+e^{2t}}\pm
\frac{\sqrt{x^2-(1+e^{2t})}}{1+e^{2t}}
\label{neseven}
\een
Expressed in terms of the coordinates $(q,\tau)$ defined in
(\ref{bfive}) these become
\ben
\bar{P}_\pm (q,\tau) = - \frac{\cosh q~e^{2\tau}}{\sqrt{1-e^{2\tau}}}
\pm \sinh q \sqrt{1-e^{2\tau}}
\label{nesevena}
\een
We will call the solution $P_+$ the upper branch of the fermi sea, and
$P_-$ the lower branch of the fermi sea.

Consider a ripple which is a perturbation of the upper branch
$\bar{P}_+(x,t)$ at early times. Since the fermi surface is itself
time dependent, at late times the ripple can appear either as a
perturbation of the lower branch or the upper branch, depending on the
initial condition. To determine which branch the final ripple ends up
in, it is sufficient to consider a point exactly on the fermi
surface. The motion of a generic point may be written as 
\ben
x(\tau_0,t)=-(\cosh\tau_0e^t+\frac{1}{2}e^\tau_0e^{-t})
\label{nefive}
\een
where $\tau_0$ parametrizes the particular point. It is clear from the
solution (\ref{neseven}) that the upper and lower branch meet at 
\ben
x=-\sqrt{1+e^{2t}}
\een
This means that if
\ben
x(\tau_0,t)<-\sqrt{1+e^{2t}} 
\label{nefiveg}
\een
for all times, 
the initial point remains in the upper branch. From (\ref{nefive})
the condition (\ref{nefiveg}) implies
\ben
\sinh^2\tau_0~e^{2t}+\frac{1}{4} e^{2\tau_0}e^{-2t}+(e^\tau_0\cosh
\tau_0 -1) > 0
\een
A sufficient condition for this to happen is
\ben
\tau_0 > 0
\een
It is straightforward to see that this is also a necessary condition,
by explicitly analyzing the trajectory equation for small $\tau_0 <0$.

To understand the significance of this condition it is useful to express
the trajectory in terms of the coordinates $(q,\tau)$ defined in (\ref{afive}).
In terms of $\tau$,  (\ref{nefive}) simply becomes
\ben
x(\tau_0,\tau) = -\frac{\cosh(\tau_0-\tau)}{\sqrt{1-e^{2\tau}}}
\een
In other words the trajectory is simply described by
\ben
q+\tau = \tau_0
\een
In other words, $\tau_0$ is the retarded time in the closed string
interpretation and gives us two distinct
situations for the scattering: for $\tau_0<0 $, the scattered pulse is
a ripple on the lower branch, while for $\tau_0>0 $ it remains on the
upper branch.  This is in exact correspondence
with the behavior of an excitation of the collective field theory at the
linearized level discussed in the previous section. 
In that case, the excitation got reflected by the mirror at $q=0$
for the retarded time $\tau_0 < 0$,
while for $\tau_0 >0$ 
the excitation never reaches the mirror. However, the behavior of
the pulse discussed in this section is {\em exact} at the classical
level and therefore includes effects of the interactions in collective
field description.

\section{The Scattering Equation}

We would like to understand how an initial small fluctiation produced
around the closing hyperbola solution background evolves in time.
Perturbing around the classical solution we define fluctuation
fields $\eta_\pm (x,t)$ as follows 
\ben
P_\pm(x,t)=\bar{P}_\pm (x,t) \pm \eta_\pm (x,t)
\een
The initial time will be taken to be localized near some large
negative value of $x = - x_1$ at an early time $t \rightarrow
-\infty$. In terms of the $q,\tau$ variables  defined in (\ref{bfive})
this means
\ben
q_1 \rightarrow \infty~~~~~\tau_1 \rightarrow -\infty
\label{cond1}
\een
with some finite value of the retarded time
\ben
q_1 + \tau_1 = \tau_0 = {\rm finite}
\label{cond2}
\een
The aim is to find the behavior of the pulse at $t = t_2 = \infty$
at some finite value of $q_2$
\ben
\tau_2 \rightarrow \infty~~~~~~~~q_2 = {\rm finite}
\label{cond3}
\een
Now let us study both cases separately:

\subsection{The case $\tau_0<0$}

For this case, the scattered pulse reaches the lower branch of the
fermi surface and the scattering equations will be
\ben
t_1 = t_2-2(t_2-\alpha) = t_2-\ln{\frac{x_1+P_-(t_2,x_1)}{x_1-P_-(t_2,x_1)}} 
\label{timedelay}
\een
where we have used (\ref{nefour}).
The condition $p(t_1) = - p(t_2)$ which describes scattering becomes
\ben
P_-(x_1,t_2)=-P_+(x_1,t_1)
\label{scattering2}
\een
Using the fluctuation fields this is
\ben
\eta_-(x_1,t_2)=\eta_+(x_1,t_1) + [\bar{P}_+(x_1,t_1)+\bar{P}_-(x_1,t_2)]
\label{neeight}
\een
A straightforward calculation using (\ref{nesevena}) yields
\ben
[\bar{P}_+(x_1,t_1)+\bar{P}_-(x_1,t_2)]
= \frac{1}{\sqrt{1-e^{2\tau_1}}}[-e^{-q_1}+e^{q_1+2\tau_1}+
\frac{\cosh q_2}{\cosh q_1}(1-e^{2\tau_1})]
\label{neeighta}
\een
In deriving this we have used the fact that the points $q_2$ and
$q_1$ refer to the same value of $x=x_1$ albeit at different times, so
that
(\ref{bfive}) yields
\ben
x_1=-\frac{\cosh q_1}{\sqrt{1-e^{2\tau_1}}}
=-\frac{\cosh q_2}{\sqrt{1-e^{2\tau_2}}}
\label{qrelation}
\een
In the limit (\ref{cond1}-\ref{cond3}), the right hand side
of (\ref{neeighta}) vanishes as $e^{\tau_1}$. Therefore the
scattering equation simply becomes
\ben \eta_-(x_1,t_2)=\eta_+(x_1,t_1)
\label{neten}
\een
It is useful to define new fields $\Phi_\pm (q,t)$ by the relations
\ben
\eta_\pm(x,t)=\frac{\sqrt{1-e^{2\tau}}}{\sinh q}\Phi_\pm(q,\tau)
\label{phidef}
\een
The motivation for introducing these factors is the following. The
perturbations $\eta_\pm$ are related to the collective field
$\varphi$ in a complicated way. However at the linearized level these
relationships become
\ben
\eta_\pm(x,t)=\frac{\sqrt{1-e^{2\tau}}}{\sinh q}(\partial_\tau \pm 
\partial_q) \varphi
\een
Since $\varphi$ is a massless field in $(q,\tau)$ space at the
linearized level, it is clear that in the same approximation
$\Phi_\pm$ are {\em chiral} fields. However, this linearized
approximation is valid only at large $q$. In our scattering problem,
the initial pulse is at $q_1 \sim \infty$ and therefore we can 
take 
\ben
\Phi_+ (q_1,\tau_1) = \Phi_+ (\tau_1+q_1)
\een
Since the final pulse is at finite $q_2$, $\Phi_-(q_2,\tau_2)$
is generally a function of both $q_2$ and $\tau_2$.

In terms of $\Phi_\pm$ the scattering equation becomes
\ben
\Phi_{out}(q_2,\tau_2)= 
\Phi_-(q_2,\tau_2) = \frac{\tanh q_2}{\tanh q_1} \Phi_+ (q_1,\tau_1)
\een
In the limits (\ref{cond1})-(\ref{cond3}) this simplifies to
\ben
\Phi_{out}(q_2,\tau_2) = \Phi_-(q_2,\tau_2) = {\tanh q_2}~\Phi_+ (\tau_1+q_1)
\label{finalscattering}
\een

We now obtain an expression for the time delay (\ref{timedelay})
when (\ref{cond1})-(\ref{cond3}) holds. Using the equations
(\ref{neseven}) and (\ref{qrelation}) repeatedly we get 
\bea
\Delta & = & \frac{x_1+P_-(x_1,t_2)}{x_1-P_-(x_1,t_2)} = 
\frac{x_1+\bar{P}_-(x_1,t_2)-\eta_-(x_1,t_2)}
{x_1-\bar{P}_-(x_1,t_2)+\eta_-(x_1,t_2)} \nn \\
& = & \frac{2\cosh^2 q_1}{(1-e^{2\tau_1}) \cosh q_2
(e^{-q_2}-\frac{\Phi_-(\tau_2,q_2)}{\sinh q_2})}-1
\eea
This expression simplifies considerably when the conditions
(\ref{cond1})-(\ref{cond3}) hold.
\ben
\ln\Delta\approx 2q_1-\ln\left[2\cosh q_2 \left( e^{-q_2}-\frac{\Phi_-(q_2,0)}{\sinh
q_2}\right)\right] 
\een 
so that the time delay in this limit becomes
\ben
t_1 = t_2 - 2q_1 + \ln \left[ 1 + e^{-2q_2}- 2\coth q_2~\Phi_-(q_2,0)\right]
\label{neeleven}
\een
In the 
limits defined in (\ref{cond1})-(\ref{cond3}) it is easy to see that
\ben
\tau_1+q_1 \sim t_1 + q_1 =  t_2 - q_1 +
\ln \left[ 1 + e^{-2q_2}- 2\coth q_2~\Phi_-(q_2,0) \right]
\label{netwelvea}
\een
In the same limit we can again use the definitions of $(q,\tau)$ and
the relation (\ref{qrelation}) to show
\ben
t_2 - q_1 \sim -\ln(2\cosh q_2)
\label{neelevena}
\een
Therfore the scattering equation (\ref{finalscattering}) becomes
\ben
\Phi_-(q_2,0) \approx
 (\tanh q_2)~\Phi_+ \left( {\ln{\frac{1+e^{-2q_2}-2\coth q_2\Phi_-(q_2,0)}{2\cosh
       q_2}}} \right)
\label{netwelve}
\een
In terms of the redefined scattered pulse
\ben
\Psi_{out}(q_2,0) = \coth~q_2~\Phi_-(q_2,0)
\label{newxi}
\een
we have the simple scattering equation
\ben
\Psi_{out}(q_2,0) \approx
\Phi_+ \left( {\ln{\frac{1+e^{-2q_2}-2\Psi_{out}(q_2,0)}{2\cosh
       q_2}}} \right)
\label{netwelvek}
\een

At this point, one could be worried that the expression inside the
logarithm in (\ref{netwelve}) can become negative for
certain values of ($q_2$,$\tau_2$). However we will show hereafter
that the condition of the existence of scattering automatically rules
out the possibility of a singular behavior. The proof of this
assertion goes as follows.  From (\ref{nefour}) we can see that as $t
\rightarrow -\infty$,  $P_+(x,t)\rightarrow \sqrt{x^2-1}$ and
therefore $P_+(x,t)  =  \bar{P}_+(x,t)+\eta_+(x,t) =
\sqrt{x^2-1}+\eta_+(x,t)$. 
The time delay equation then implies
\ben
e^{t_1-t_2}=\frac{y-\sqrt{y^2-1}-\eta_+}{y+\sqrt{y^2-1}+\eta_+} 
\een
where we have defined the positive quantity $y \equiv - x_1$.
Since the left hand side of this equation is always a positive
quantity, consistency requires
\ben
y-\sqrt{y^2-1}-\eta_+>0 
\een 
Using the definition of $\Phi_+$ in (\ref{phidef}) and the fact
that at $\tau_1 \rightarrow -\infty$, (\ref{qrelation}) implies
$x_1 = -\cosh q_1$ this becomes
\ben 
\cosh q_1-\sinh q_1-\frac{1}{\sinh q_1}\Phi_+>0 
\een 
which implies (neglecting terms
of order $e^{-2q}$) 
\ben 
\Phi_+<\frac{1}{2} 
\label{ggone}
\een 
The basic scattering equation (\ref{finalscattering}) then implies
\ben 
2\coth q_2 \Phi_-< 1 
\een
This relation implies that \ben 1+e^{-2q_2}-2\coth q_2\Phi_->0 \een
which proves the consistency of our scattering equation.

For $q_2 \gg 0$ the scattering equation (\ref{netwelve}) becomes
\ben
\Phi_-(q_2,0) \approx
 \Phi_+ \left( -q_2 + \ln[1-2\Phi_-(q_2,0)] \right)
\een
To lowest order in $\Phi_-$, the scattered pulse is therefore
peaked at the value of $|\tau_0|$, exactly as in 
linearized collective field theory. This is expected since both the
incident and the scattered pulses are in the weak coupling region.

For $q_2 \sim 0$, the quantity $\Psi_{out}$ satisfies to lowest order
\ben
\Psi_{out}(q_2,0) = \Phi_+ \left( \ln [1-\Psi_{out}(q_2,0)] \right)
\een

\subsection{The case $\tau_0>0$}

For $\tau_0 > 0$ the final pulse at $t_2 \rightarrow \infty$ or
$\tau_2 \rightarrow 0$ remains on the upper branch of the fermi
surface. This implies a modification of the various formulae in the
previous subsection. Basically we have to make the replacement
\ben
P_-(x_1,t_2) \rightarrow P_+(x_1,t_2)~~~~~~~~~~~~~~
\eta_-(x_1,t_2) \rightarrow -\eta_+(x_1,t_2)
\een

In particular, the time delay equation (\ref{timedelay}) gets modified to
\ben
t_1 = t_2-2(t_2-\alpha) = t_2-\ln{\frac{x_1+P_+(t_2,x_1)}{x_1-P_+(t_2,x_1)}} 
\label{timedelay2}
\een
while the basic scattering equation becomes
\ben
P_+(x_1,t_2)=-P_+(x_1,t_1)
\label{scattering3}
\een
In a way entirely analogous to the derivation of (\ref{neeighta}), the
quantity $[\bar{P}_+(x_1,t_1)+\bar{P}_+(x_1,t_2)]$ now vanishes
exponentially fast so that the scattering equation reduces to 
\ben 
\eta_+(x_1,t_2)=-\eta_+(x_1,t_1)
\label{neten2}
\een
The remaining steps to the final scattering equation are also
identical, leading to
\ben
\Phi_{out}(q_2,0) \equiv
\Phi_+(\tau_2,q_2) = - {\tanh q_2}~\Phi_+ (\tau_1+q_1)
\label{finalscattering2}
\een
Once again in the limit (\ref{cond1})-(\ref{cond3}) the expression for
the time delay simplifies, which now leads to, instead of
(\ref{netwelvea})
\ben
\tau_1+q_1 \sim t_1 + q_1 =  t_2 - q_1 +
\ln [1 + e^{2q_2}- 2\coth q_2~\Phi_+(\tau_2,q_2)]
\label{netwelveb}
\een
Using (\ref{neelevena}) the final equation which yields the
scattered pulse at late times becomes, instead of (\ref{netwelve})
\ben
\Phi_{out}(q_2,0) = \Phi_+(q_2,0) \approx
- (\tanh q_2)~\Phi_+ \left(  {\ln{\frac{1+e^{2q_2}-2\coth q_2\Phi_+}{2\cosh
       q_2}}} \right)
\label{netwelvec}
\een

\section{Behavior of the Scattered pulse}

In this section we numerically investigate the behavior of the
scattered pulse for a given initial pulse for various values of
$\tau_0$. We start with a gaussian pulse which is centered at 
a retarded time equal to $\tau_0$, i.e. the function $\Phi_+$ is
\ben
\Phi_+(w) = A~{\rm exp}~[-\frac{(w-\tau_0)^2}{a^2}]
\een
\begin{figure}[ht]
\centerline{\epsfxsize=1.5in
\epsfysize=2.0in
   {\epsffile{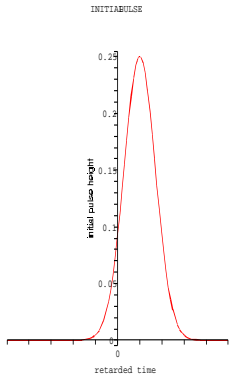}}
}
\caption{The initial pulse, i.e. $\Phi_+$ as a function of the
retarded time}
\label{initialpulse}
\end{figure}
with some constant
$A$ which is small, so that the condition (\ref{ggone}) is satisfied.
We then numerically find the final scattered
pulse at $t \rightarrow \infty$ or equivalently $\tau = 0$ 
using (\ref{netwelve}) or (\ref{netwelvec}).

\begin{figure}[ht]
\centerline{\epsfxsize=1.5in
\epsfysize=2.0in
   {\epsffile{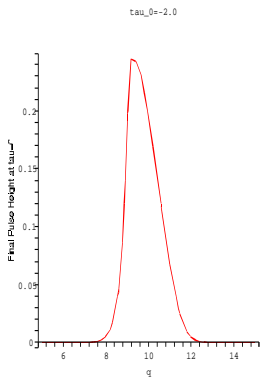}}
}
\caption{The scattered pulse for $\tau_0 = -2$}

\label{tauminus2}
\end{figure}
\begin{figure}[ht]
\centerline{\epsfxsize=1.5in
\epsfysize=2.0in
   {\epsffile{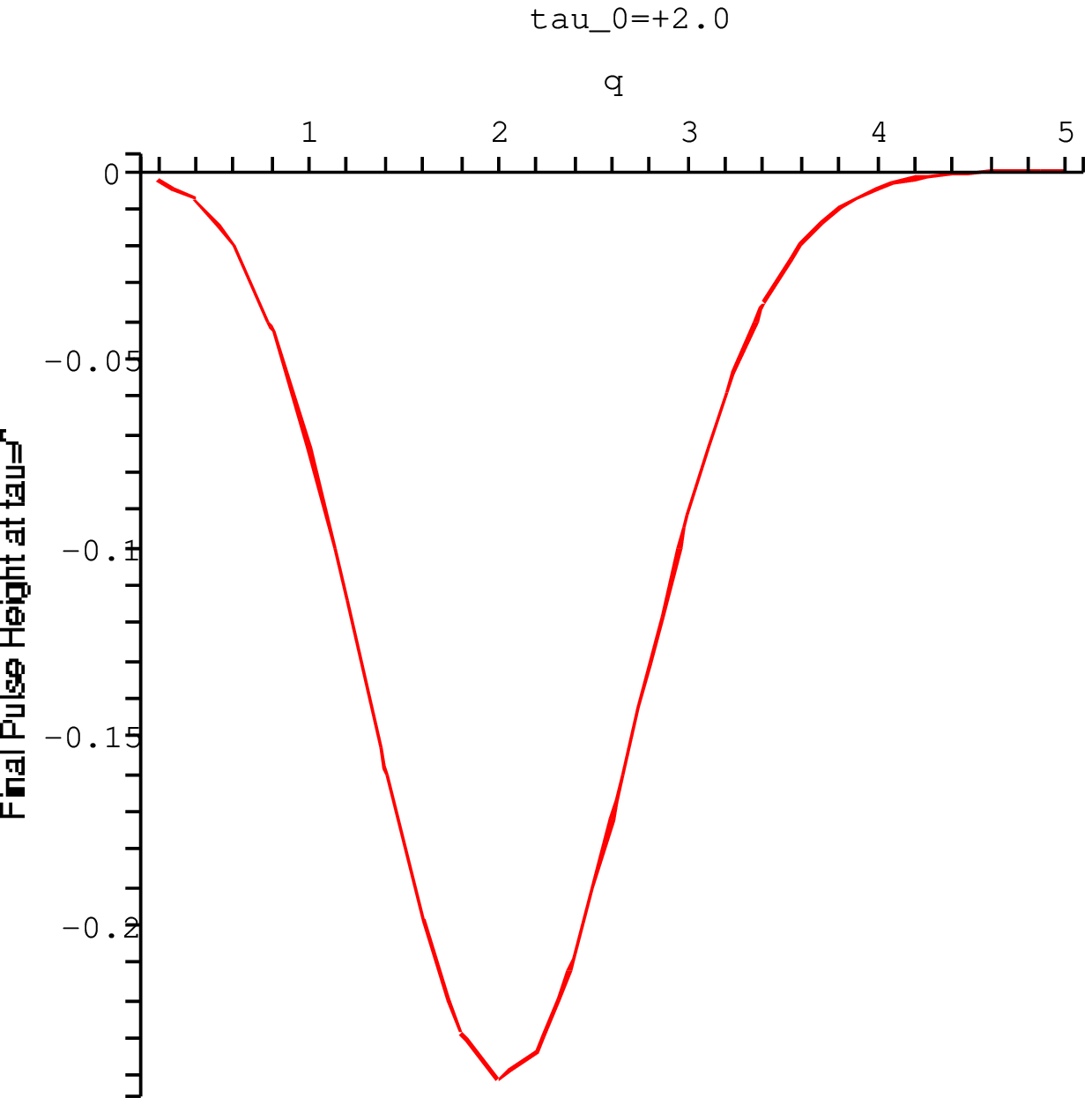}}
}
\caption{The scattered pulse for $\tau_0 = 2$}

\label{tauplus2}
\end{figure}

At the level of linearized collective field theory, the
behavior of the scattered pulse is described at the end of section 2
and depicted in Figure 3. This shows that at this level of
approximation, the scattered pulse is also a gaussian which is peaked
at $q = |\tau_0|$. For $\tau_0 < 0$ this happens due to a reflection
from the mirror. Such a reflection would invert the pulse - however we
have defined the quantity $\Phi_-$ above to incorporate this.
For $\tau_0 > 0$ there is no reflection - this means the $\Phi_{out}$ now
has the opposite sign of $\Phi_+$.

This behavior is exactly what we observe in the exact solution.
Figure (\ref{initialpulse}) shows the initial pulse, while the
Figures (\ref{tauminus2}) and (\ref{tauplus2}) are the scattered
pulses for $\tau_0 = -2$ and $\tau_0 = 2$ respectively. These are
reasonably large values of $|\tau_0|$. Note the scattered pulse is
also centered around $q=-2$ and $q=2$ respectively. This is expected
since for large values of $q$ on the $\tau=0$ surface, the collective
field theory is reasonably weakly coupled and the linearized
approximation reliable.
Furthermore, as we would expect there is not much deformation of the
pulse.

\begin{figure}[ht]
\centerline{\epsfxsize=1.5in
\epsfysize=2.0in
   {\epsffile{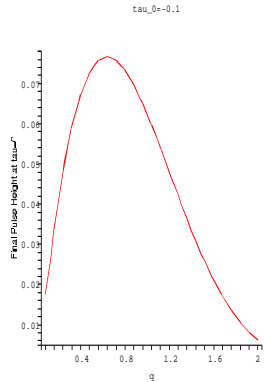}}
}
\caption{The scattered pulse for $\tau_0 = -0.1$}

\label{tauminus01}
\end{figure}

\begin{figure}[ht]
\centerline{\epsfxsize=1.5in
\epsfysize=2.0in
   {\epsffile{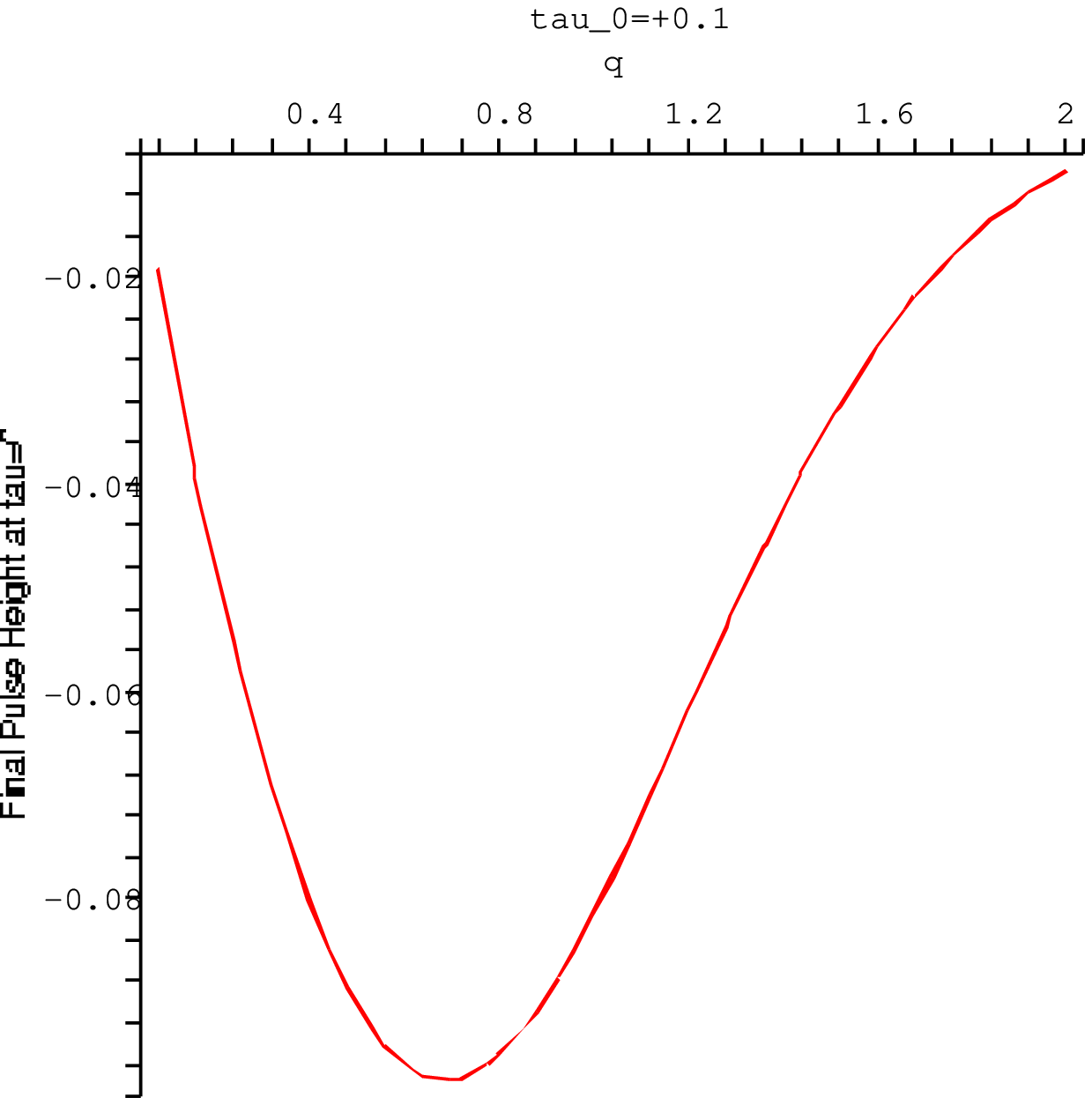}}
}
\caption{The scattered pulse for $\tau_0 = 0.1$}

\label{tauplus01}
\end{figure}

Figures (\ref{tauminus01}) and (\ref{tauplus01}) show the scattered
pulse for {\em small} values of $|\tau_0|$. 

Unlike the previous cases, the scattered pulses are centered 
at values of $q > |\tau_0|$. This is then the effect of nonlinearities
in the collective field description which are expected to be strong at
small values of $q$. This trend becomes more pronounced as we go to
smaller values of $|\tau_0|$

The main point is that the scattering problem is well defined for all
values of $\tau_0$ and the scattered pulse is smooth, always vanishing
at $q=0$. In the original space $x$ provided by the eigenvalues of the
matrix, the background fermi sea gets completely drained out and
ceases to exist for any finite $x$. It might appear from this that the
scattering problem is pathological since the pulses do not seem to
have any ``space'' to move in at infinitely late times. Our discussion
shows that this is not the ``space'' in which the scattering problem
has to be formulated. In the closed string space, $q$, scattering
makes perfect sense and smooth. While the closed string time ends at
$\tau=0$ the real time evolution of the ripple on the fermi sea is
over a complete time range, and the fact that ${\cal I}^+$ is not
weakly coupled does not pose any problem for the scattering data.

\section{Fermion correlators in time dependent backgrounds}

In this section, we will outline a method to obtain {\em exact} fermionic
correlators in the nontrivial time dependent background.
We will calculate the two point function explicitly and show that there
is no pathological behavior at late times. This is something we already know
at the classical level from the previous section. However the following
considerations take into account all quantum corrections.

The strategy is to use the fact that these time dependent
solutions are obtained from the ground state by the action of 
non-diagonal $W_\infty$ transformations, as described by the equation
(\ref{vthree}). As discussed extensively in
\cite{Das:1991qb, Dhar:1992rs}, these symmetry generators can be
obatained from the basic quantity
\ben
W(\alpha,\beta,t)=\frac{1}{2} \int dx~e^{i\alpha x}
\psi^\dag(x+\beta/2,t)
\psi(x-\beta/2,t)
\label{cortwo}
\een
In terms of $W(\alpha,\beta,t)$, the charges are $W_{r,s}$ :
\ben
W_{rs} = e^{-(r-s)t}\int d\alpha d\beta~ F_{rs}
\left(\frac{\alpha-\beta}{\sqrt2},\frac{\alpha+\beta}{\sqrt2}\right) W(\alpha,\beta,t)
\label{corthree}
\een
where
\ben
F_{rs}(a,b)=2\cos\left(\frac{ab}{2}\right)(i\partial_a)^r 
(i\partial_b)^s\delta(a)\delta(b)
\label{corfour}
\een
The basic quantity in the fermionic theory is the Wigner operator
\ben
u(x,p) \equiv \int dye^{ipy} : \psi^\dag(x-y/2)\psi(x+y/2) :
\label{corfive}
\een
Under the $W_\infty$ transformations this quantity changes by the
basic commutator \cite{Dhar:1992rs}
\ben
[W(\alpha,\beta,t),u(x,p)]=\frac{1}{2}e^{i(\alpha x-\beta p)}
\left( u(x-\beta/2,p-\alpha/2)-u(x+\beta/2,p+\alpha/2)\right)
\label{corsix}
\een
Using these expressions, we can calculate the change of the
phase space density operator $u(x,p,.t)$ under a finite transformation
\ben
u(x,p) \rightarrow u^\prime (x,p) = \exp(-i\lambda W_{rs})u(x,p,t)\exp(i\lambda
W_{rs})
\label{corseven}
\een
Now consider a time dependent background which is described by a
non-normalizable state
\ben
|\lambda> = e^{i\lambda W_{rs}}|\mu>
\label{coreight}
\een
where $|\mu>$ denotes the ground state. 
Then we have the identity
\ben
<\lambda | u(x_1,p_1) u(x_2,p_2) \cdots  |\lambda>
= < \mu | u^\prime (x_1,p_1)~u^\prime (x_2,p_2) \cdots |\mu>
\label{corone}
\een 
Since the operator
\ben
: \psi^\dagger (x_1) \psi (x_2) :
\een
can be expressed in terms of the operator $u(x,p)$ we can use
(\ref{corone}) to calculate correlators like
\ben
<\lambda| : \psi^\dagger (x_1) \psi (y_1) : : \psi^\dagger (x_2) \psi
(y_2) : \cdots |\lambda> 
\een
in terms of the correlators in the ground state. The latter have been
calculated eaxctly in \cite{Moore:1991sf}, which may be then used
to calculate the correlators in the nontrivial background.

Essentially the same philosophy was used in 
\cite{Karczmarek:2004ph} for the draining fermi sea solution.
However in that case, the corresponding $W_\infty$ transformation did
not involve a nontrivial transformation of the momentum variable. This
meant that the correlators of density operators $\psi^\dagger
\psi$ in the nontrivial
background can be written 
in terms of correlators of density operators in the ground state. In
the cases of our interest, we need correlators of higher moments of the
phase space density.

In the following we will perform an explicit computation of the expectation
value of the density operator in the closing hyperbola solution.

\subsection{Density operator in the Closing Hyperbola solution}

The density operator $\rho$ is
\ben
\rho (x,t) = : \psi^\dag(x,t)\psi(x,t) = \frac{1}{2\pi}\int dp~ u(x,p,t)
\label{coreleven} 
\een
The state in the fermionic theory which corresponds to the closing hyperbola
solution is 
\ben
|\lambda> = e^{i\lambda W_{02}}|\mu>
\label{cortwelve}
\een
For $r=0,s=2$ the relations (\ref{corthree}) and (\ref{corfour}) become
\ben
W_{02}=-e^{2t}(\partial_\alpha+\partial_\beta)^2 W(\alpha,\beta,t) |_{\alpha=
\beta = 0}
\label{eqone}
\een
A straightforward calculation using (\ref{corsix}) then yields
\ben
[u(x,p),W_{02}]=e^{2t}(-i)(x-p)(\partial_x+\partial_p)u(x,p)
\label{cornine}
\een
The form of this commutator immediately shows that a {\em finite} transformation of
the Wigner operator itself is 
\ben
\exp(-i\lambda W_{02})u(x,p)\exp(i\lambda W_{02})=
e^{\lambda e^{2t}(x-p)(\partial_x+\partial_p)}~u(x,p)
= u(x^\prime, p^\prime)
\label{corten}
\een
where we have defined
\ben
x^\prime = x + \lambda e^{2t} (x-p)~~~~~~~p^\prime = p + \lambda e^{2t} (x-p)
\label{corthirteen}
\een
This is what one would expect at the classical level since these are precisely
the transformations in the single particle phase space. What we have shown, however,
is that the result is exact at the quantum level.

The above results show that the expectation value of the density operator 
in the closing hyperbola state $|\lambda>$ is given in terms of the two point
fermion correlator in the ground state $|\mu>$ by
\ben
<\lambda | \rho (x) | \lambda >
= \frac{1}{2\pi}\int dy~dp~e^{ip^\prime y}
<\mu | \psi^\dagger (x^\prime - \frac{y}{2}) \psi (x^\prime + \frac{y}{2})
| \mu >
\label{cprfourteen}
\een
with $x^\prime, p^\prime$ given by (\ref{corthirteen}).

In \cite{Moore:1991sf} the two-point fermion correlator was determined 
in the ground state as \footnote{In \cite{Moore:1991sf} the single particle hamiltonian
is taken as $h = p^2 - \frac{1}{4}x^2$ which differs from ours by rescaling of
$p$ and $x$. The formula (\ref{eqthree}) differs from that in
\cite{Moore:1991sf}) since takes into account this rescaling as well as
a rescaling of the fermion field necessary to preserve the correct anticommutation
relation.}
\ben
<\psi\dag(x_1)\psi(x_2)>_\mu = 
\sqrt2 i\int_{-\infty}^{+\infty}\frac{dq}{2\pi}\int_0^{sgn(q)\infty} 
ds \frac{e^{-sq+is\mu}}{(-4\pi i \sinh s)^{1/2}}
e^{\cH(x_1,x_2)}
\label{eqthree}
\een
where
\ben
\cH(x_1,x_2)=-\frac{i}{2}\left(\frac{x_1^2+x_2^2}
{\tanh s}-2\frac{x_1x_2}{\sinh s}\right)
\een
Using (\ref{corthirteen}) we therefore get
\ben
<\lambda | \rho (x,t) | \lambda > = \sqrt2 \frac{1}{2\pi} i\int_{-\infty}^{+\infty}dp
\int_{-\infty}^{+\infty}dye^{ip'y} i \int_{-\infty}^{+\infty}
\frac{dq}{2\pi}\int_{0}^{sgnq(\infty)} ds ~F(s)
~G(s,x^\prime,y) + (s\rightarrow -s)
\een
where
\ben
F(s)=\frac{e^{-sq+is\mu}}{(-4i\pi\sinh s)^{1/2}}~~~~~~~
G(s,x^\prime,y) = \exp[-\frac{i}{2}(2\tanh\frac{s}{2}x'^2+\coth\frac{s}{2}\frac{y^2}{2})]
\een
Performing the integration over $y$ we get
\ben
<\lambda | \rho (x,t) | \lambda > = \frac{\sqrt2}{2\pi}[i  \int\frac{ds}{2\pi s}
(\frac{-4\pi i}{\coth s/2})^{1/2}
\frac{e^{is\mu}}{(-4i\pi\sinh s)^{1/2}}\int_{-\infty}^{+\infty} 
dp e^{-i\tanh\frac{s}{2}(x'^2-p'^2)} +c.c]
\label{eqfour}
\een
Defining the functions
\ben
f(t) = 1 + 2\lambda e^{2t} ~~~~h (t) = -2\lambda e^{2t}~~~~~ g(t) = -(1-2\lambda e^{2t})
\een
(\ref{corthirteen}) yields
\ben
x'^2-p'^2 = -\frac{1}{g(t)}x^2 + g(p + \frac{h(t) x}{g(t) })^2
\label{eqeight}
\een
so that intgerating over $p$ we get the final answer
\ben
<\lambda | \rho (x,t) | \lambda > = 2\sqrt2 [i(\frac{-1}{g(t)})^{1/2} 
\int _{0}^{\infty}\frac{ds}{2\pi s}
\frac{e^{is\mu-\frac{i\tanh(s/2)x^2}{-g (t) }}}{(-4i\pi\sinh s)^{1/2}} +c.c ]
\label{eqnine}
\een
The density expectation value therefore satisfies the remarkably simple relation
\ben
<\lambda | \rho (x,t) | \lambda > = \frac{1}{(1-2\lambda e^{2t})^{1/2}}
<\mu | \rho ( \frac{x}{{\sqrt{1-2\lambda e^{2t}}}},t) | \mu > 
\label{cofifteen}
\een

It is easy to check that this exact expression leads to the correct
semiclassical answer. To do this, it is necessary to perform
a derivative with respect to $\mu$,
\ben
\partial_{\mu}<\lambda | \rho (x,t) | \lambda >  = 
-2\sqrt2 [(\frac{-1}{g(t)})^{1/2} \int _{0}^{\infty}\frac{ds}{2\pi }
\frac{e^{is\mu-\frac{i\tanh(s/2)x^2}{-g(t)}}}{(-4i\pi\sinh s)^{1/2}} +c.c ]
\label{eqten}
\een
and consider the limit $s \approx 0$ in (\ref{eqten}). 
This leads, after redefining $z^2 = \frac{x^2}{-g(t)}$,  to:
\ben
\partial_{\mu}<\lambda | \rho (x,t) | \lambda > = -2\sqrt2 [(\frac{-1}{g})^{1/2} 
\int _{0}^{\infty}\frac{ds}{2\pi }\frac{1}{(-4\pi i)^{1/2}}
\frac{i(\mu -\frac{z^2}{2})s}{\sqrt s} +c.c ]
\een
Integration over $s$ then gives
\ben
\partial_{\mu} <\lambda | \rho (x,t) | \lambda > = 
-\frac{1}{(-g)^{1/2}} \frac{1}{\pi \sqrt {z^2 - 2\mu}}
\een
To compare with the semiclassical answer (\ref{afive}) for the closing
hyperbola, we need to set
$2\mu = 1$ and $2\lambda = -1$ \footnote{Note that the $\lambda$ of this
section is related to $\lambda_+$ of Section 2.2 by 
$2\lambda = \lambda_+$}. This yields
\ben
<-\frac{1}{2}| \rho (x,t) |-\frac{1}{2} > = \frac{1}{\pi}
\frac{\sqrt {x^2 - (1+e^{2t})} }{1+e^{2t}}
\een
which is identical to (\ref{afive}).

The exact answer in fact corroborates our conclusions about the nature of
the background based on the semiclassical solution. At large times, equation
(\ref{cofifteen}) shows that the density goes to zero at any finite $x$. However,
as our previous sections show the physics at late times occurs at infinite values
of $|x|$ and finite values of the closed string coordinate $q$.

Our result for the one point function of the eigenvalue density does not
reveal any pathological behavior. The scattering matrix is related to higher
point functions which may be calculated using similar techniques. It is 
important to see whether these have any interesting behavior coming from
nonperturbative effects.

\section{Acknowledgements} We would like to thank Joanna Karczmarek
for many discussions and collaboration at early stages of this work, 
and Lenny Susskind for a correspondence. This work is supported in 
part
by National Science Foundation grants Nos. PHY-0244811 and PHY-0555444
and by Department of Energy contracts No. DE-FG01-00ER45832.
and No. DE-FG02-91ER-40690.


\begin{thebibliography}{99}

\bibitem{Das:2004aq}
  S.~R.~Das and J.~L.~Karczmarek,
  Phys.\ Rev.\ D {\bf 71}, 086006 (2005)
  [arXiv:hep-th/0412093]; S.~R.~Das, 
    Mod.\ Phys.\ Lett.\ A {\bf 20}, 2101 (2005). 

\bibitem{Craps:2005wd}
  B.~Craps, S.~Sethi and E.~P.~Verlinde,
  JHEP {\bf 0510}, 005 (2005)
  [arXiv:hep-th/0506180];
  B.~Craps, A.~Rajaraman and S.~Sethi,
  Phys.\ Rev.\ D {\bf 73}, 106005 (2006)
  [arXiv:hep-th/0601062];
  B.~Craps,
  Class.\ Quant.\ Grav.\  {\bf 23}, S849 (2006)
  [arXiv:hep-th/0605199].

\bibitem{li0506}
M.~Li,
Phys.\ Lett.\ B {\bf 626\/}, 202--208 (2005)
[hep-th/0506260];M.~Li and W.~Song,
JHEP {\bf 10\/}, 073 (2005)
[hep-th/0507185];B.~Chen,
  Phys.\ Lett.\ B {\bf 632\/}, 393--398 (2006)
  [hep-th/0508191]; J.~H.~She,
JHEP {\bf 01\/}, 002 (2006)
[hep-th/0509067]; M.~Li and W.~Song,
JHEP {\bf 08\/}, 089 (2006)
  [hep-th/0512335]; H.~Chen and B.~Chen,
Phys.\ Lett.\  B {\bf 638\/}, 74--79 (2006)
[hep-th/0603147]; T.~Ishino and N.~Ohta,
time-dependent solutions,''
Phys.\ Lett.\ B {\bf 638\/}, 105--109 (2006)
[hep-th/0603215];
  H.~Kodama and N.~Ohta,
  Prog.\ Theor.\ Phys.\  {\bf 116}, 295 (2006)
  [arXiv:hep-th/0605179];
  T.~Ishino, H.~Kodama and N.~Ohta,
  Phys.\ Lett.\  B {\bf 631}, 68 (2005)
  [arXiv:hep-th/0509173].

\bibitem{Das:2005vd}
  S.~R.~Das and J.~Michelson,
  Phys.\ Rev.\ D {\bf 72}, 086005 (2005)
  [arXiv:hep-th/0508068]; 
  S.~R.~Das and J.~Michelson,
  Phys.\ Rev.\ D {\bf 73}, 126006 (2006)
  [arXiv:hep-th/0602099].

\bibitem{robbinssethi0509}
D.~Robbins and S.~Sethi,
JHEP {\bf 02\/}, 052 (2006)
[hep-th/0509204]; D.~Robbins, E.~Martinec and S.~Sethi,
JHEP {\bf 08\/}, 025 (2006)
[hep-th/0603104].

\bibitem{Das:2006pw}
S.~R.~Das, J.~Michelson, K.~Narayan and S.~P.~Trivedi,
  Phys.\ Rev.\  D {\bf 75}, 026002 (2007)
  [arXiv:hep-th/0610053];
  S.~R.~Das, J.~Michelson, K.~Narayan and S.~P.~Trivedi,
  Phys.\ Rev.\ D {\bf 74}, 026002 (2006)
  [arXiv:hep-th/0602107].

\bibitem{Chu:2006pa}
  C.~S.~Chu and P.~M.~Ho,
  JHEP {\bf 0604}, 013 (2006)
  [arXiv:hep-th/0602054].

\bibitem{refer} A partial list of references may 
be found in \cite{Das:2006pw}.
 
\bibitem{McGreevy:2003kb}
J.~McGreevy and H.~Verlinde,
JHEP {\bf 0312}, 054 (2003)
[arXiv:hep-th/0304224];
I.~R.~Klebanov, J.~Maldacena and N.~Seiberg,
JHEP {\bf 0307}, 045 (2003)
[arXiv:hep-th/0305159];
A.~Sen,
Mod.\ Phys.\ Lett.\ A {\bf 19}, 841 (2004)
[arXiv:hep-th/0308068].


\bibitem{Jevicki:1979mb}
A.~Jevicki and B.~Sakita,
Limit,''
Nucl.\ Phys.\ B {\bf 165}, 511 (1980).

\bibitem{Das:1990ka}
S.~R.~Das and A.~Jevicki,
Mod.\ Phys.\ Lett.\ A {\bf 5}, 1639 (1990).



\bibitem{Minic:1991rk}
D.~Minic, J.~Polchinski and Z.~Yang,
Nucl.\ Phys.\ B {\bf 369}, 324 (1992);
G.~W.~Moore and R.~Plesser,
Phys.\ Rev.\ D {\bf 46}, 1730 (1992)
[arXiv:hep-th/9203060];
S.~Y.~Alexandrov, V.~A.~Kazakov and I.~K.~Kostov,
Nucl.\ Phys.\ B {\bf 640}, 119 (2002)
[arXiv:hep-th/0205079].

\bibitem{Karczmarek:2003pv}
J.~L.~Karczmarek and A.~Strominger,
JHEP {\bf 0404}, 055 (2004)
[arXiv:hep-th/0309138]


\bibitem{Karczmarek:2004ph}
J.~L.~Karczmarek and A.~Strominger,
JHEP {\bf 0405}, 062 (2004)
[arXiv:hep-th/0403169].

\bibitem{Das:2004hw}
S.~R.~Das, J.~L.~Davis, F.~Larsen and P.~Mukhopadhyay,
Phys.\ Rev.\ D {\bf 70}, 044017 (2004)
[arXiv:hep-th/0403275].

\bibitem{Polchinski:1991uq}
  J.~Polchinski,
  Nucl.\ Phys.\ B {\bf 362}, 125 (1991).

\bibitem{Natsuume:1994sp}
  M.~Natsuume and J.~Polchinski,
  Nucl.\ Phys.\ B {\bf 424}, 137 (1994)
  [arXiv:hep-th/9402156].


\bibitem{Moore:1991sf}
  G.~W.~Moore,
  Nucl.\ Phys.\ B {\bf 368}, 557 (1992).

\bibitem{ceqone}
For reviews and references to the original literature
see e.g.I.~R.~Klebanov,
arXiv:hep-th/9108019;
S.~R.~Das,
arXiv:hep-th/9211085;
A.~Jevicki,
arXiv:hep-th/9309115;
P.~H.~Ginsparg and G.~W.~Moore,
arXiv:hep-th/9304011;
J.~Polchinski,
arXiv:hep-th/9411028;
E.~J.~Martinec,
arXiv:hep-th/0410136.

\bibitem{Dhar:1992cs}
A.~Dhar, G.~Mandal and S.~R.~Wadia,
Int.\ J.\ Mod.\ Phys.\ A {\bf 8}, 3811 (1993)
[arXiv:hep-th/9212027];


\bibitem{Das:1995gd}
S.~R.~Das and S.~D.~Mathur,
Phys.\ Lett.\ B {\bf 365}, 79 (1996)
[arXiv:hep-th/9507141];
S.~R.~Das,
arXiv:hep-th/0401067.



\bibitem{Takayanagi:2003sm}
T.~Takayanagi and N.~Toumbas,
JHEP {\bf 0307}, 064 (2003)
[arXiv:hep-th/0307083];
M.~R.~Douglas, I.~R.~Klebanov, D.~Kutasov,
J.~Maldacena, E.~Martinec and N.~Seiberg,
arXiv:hep-th/0307195.

\bibitem{Das:1991qb}
  S.~R.~Das, A.~Dhar, G.~Mandal and S.~R.~Wadia,
  Int.\ J.\ Mod.\ Phys.\  A {\bf 7}, 5165 (1992)
  [arXiv:hep-th/9110021];
  S.~R.~Das, A.~Dhar, G.~Mandal and S.~R.~Wadia,
  Mod.\ Phys.\ Lett.\  A {\bf 7}, 71 (1992)
  [arXiv:hep-th/9111021];
  S.~R.~Das, A.~Dhar, G.~Mandal and S.~R.~Wadia,
  Mod.\ Phys.\ Lett.\  A {\bf 7}, 937 (1992)
  [Erratum-ibid.\  A {\bf 7}, 2245 (1992)]
  [arXiv:hep-th/9112052].

\bibitem{Dhar:1992rs}
  A.~Dhar, G.~Mandal and S.~R.~Wadia,
  Int.\ J.\ Mod.\ Phys.\  A {\bf 8}, 325 (1993)
  [arXiv:hep-th/9204028]; 
  A.~Dhar, G.~Mandal and S.~R.~Wadia,
  Mod.\ Phys.\ Lett.\  A {\bf 7}, 3129 (1992)
  [arXiv:hep-th/9207011];
  A.~Dhar, G.~Mandal and S.~R.~Wadia,
  Mod.\ Phys.\ Lett.\  A {\bf 8}, 3557 (1993)
  [arXiv:hep-th/9309028].








\end{thebibliography}
\end{document}